Stanford | ENGINEERING
Civil & Environmental Engineering

# CIFE
CENTER FOR INTEGRATED FACILITY ENGINEERING

# Measuring the Impact of Blockchain and Smart Contract on Construction Supply Chain Visibility

By

# Hesam Hamledari & Martin Fischer

# Measuring the Impact of Blockchain and Smart Contract on Construction Supply Chain Visibility

## Hesam Hamledari[1], Martin Fischer[2]

[1] PhD Candidate, Stanford University, Department of Civil & Environmental Engineering, Stanford, CA, United States, hesamh@stanford.edu

[2] Kumagai Professor of Engineering & Professor in Civil & Environmental Engineering, Stanford University, Stanford, CA, United States, fischer@stanford.edu

## Abstract

This work assesses the impact of blockchain and smart contract on the visibility of construction supply chain and in the context of payments (intersection of cash and product flows). It uses comparative empirical experiments (Charrette Test Method) to draw comparisons between the visibility of state-of-practice and blockchain-enabled payment systems in a commercial construction project. Comparisons were drawn across four levels of granularity. The findings are twofold: 1) blockchain improved information completeness and information accuracy respectively by an average 216% and 261% compared with the digital state-of-practice solution. The improvements were significantly more pronounced for inquiries that had higher product, trade, and temporal granularity; 2) blockchain-enabled solution was robust in the face of increased granularity, while the conventional solution experienced 50% and 66.7% decline respectively in completeness and accuracy of information. The paper concludes with a discussion of mechanisms contributing to visibility and technology adoption based on business objectives.

# 1. Introduction

Increasing the visibility of the construction supply chain has long been one of the key objectives of supply chain management practices [1]. Supply chain partners need a shared understanding with respect to the flow of information, cash, and products; timely and accurate access to this data is key to achieving situational awareness [2] and one of the building blocks of successful project delivery [3]. This is particularly a challenge in the construction phase due to low transparency, fragmentation, limited communication, and inadequate information exchange between stakeholders such as clients, general contractors, and subcontractors [4].

This lack of transparency is a weak link in the construction supply chain [5,6], and it has dire implications for the industry. It creates information asymmetry between project stakeholders, a major risk factor [7] and a source of distrust [8]. It leads to poor process orientation [9] and poor coordination between project members [10]. Engineers and architects spend more than 54% of their time managing information [11], therefore the inefficiencies caused by poor information visibility directly hurt business performance of projects [12].

The blockchain technology has a *potential* to address these limitations; it is expected to enhance supply chain visibility [13] and eliminate siloed data developments [14] due to its innovative approach toward decentralized consensus. It allows for secure data sharing and collaboration in trust-less environments, creating "verifiable confidence" [15] in the data and computational systems used in construction and engineering projects. These *potentials* have led to a positive outlook from the research community, with a rapidly growing body of literature on the potential impact of blockchain and smart contracts for managing information, cash, and product flows in the construction supply chain.

Despite this optimism, the impact of blockchain and smart contracts on the visibility of the construction supply chain remains to be measured in real-world applications. The research landscape has remained mostly theoretical [16-18] and needs yet to validate the effectiveness of the technology in real-world implementations [19]. It is therefore not surprising that industry practitioners remain skeptical about the technology's potential and wary of its risks [20]. This technology will not see major industry adoption until its impact is clearly quantified [21] and contextualized with respect to current practices.

To address this limitation, the paper assesses the impact of the blockchain and smart contracts on the construction supply chain visibility and in the context of payments (intersection of cash and product flows). This work uses the Charrette Test Method to compare the state-of-the-practice payment system, used by today's industry practitioners, with a payment system based on blockchain-enabled smart contracts. The payments on a commercial construction project were processed and visualized using the two systems and for a period of three months; the two systems were contrasted in terms of the visibility of payment information at four levels of granularity. The tests provide insights with regard to 1) the impact of blockchain and smart contract on the visibility of construction supply chain; 2) the impact of increased granularity on the visibility of both conventional and blockchain-enabled solutions; and 3) how firms can adopt these technologies based on their business objectives.

The paper first discusses the underlying problems contributing to low visibility and reviews the relevant work on blockchain applications that support supply chain visibility (section 2). The sections 3, 4, and 5 respectively delve into the two research questions explored in this work, experiments designed to address the research questions, and the results of the experiments. The paper is concluded with a discussion of limitations and future work (section 6 and 7).

## 2. Point of Departure

The construction and engineering industry continues to strive toward supply chain visibility (SCV). The need for enhanced data sharing in the construction industry's multidisciplinary environment motivated efforts on computerized product modeling [22] giving rise to building information modeling (BIM) [23,24]. The application of BIM promised benefits which among others included transparency via integrated information [25] and the creation of a single source of truth. While BIM continues to transform the industry, the problem of low SCV persists [26], and the industry struggles to leverage technology for communicating information [27]. This is despite the technological advances in virtual modeling and the introduction of data exchange standards such as industry foundation classes (IFC) [28].

The industry's current information systems "do not support the duality of *product* and *project* management." [29]. The application of BIM provided a means of *product communication* [30], addressing the former objective. However, construction projects are made of social processes [31]; the management of projects and the dynamic collaborations introduce complexities that necessitate adaptations to the organization and the delivery of projects [32]. The mere availability of information does not enhance SCV [33].

It is necessary for the research efforts to go beyond product modeling and to create a collaborative environment for supply chain partners [34]. The individuals who benefit from a piece of information are typically not those who record it [27,35] nor those developing the underlying information system [36]. These independent stakeholders have competing business objectives which drives them into "information siloes" [37], hurting the integration and visibility of supply chain flows [38].

At its core, therefore, the problem of low SCV stems from a lack of trust between supply chain partners [39,40]. This problem is further exacerbated due to the industry's reliance on third-party financial institutions for managing cash flow [41-43] and the high degree of fragmentation [44,45]. To achieve SCV, it is necessary to create trust in the integrity of protocols used for data creation (information, cash, and prodcut flows) [46] and those used for data sharing [47].

These are the very limitations that the novel governance models at the heart of blockchain and smart contracts aim to address, making these technologies relevant to SCV research. The following section reviews the relevant points of departure: the studies on the application of blockchain and smart contract for enhancing the transparency of supply chain flows in the construction and engineering industry.

### *2.1. Blockchain and Smart Contracts for Enhancing Supply Chain Visibility*

Increased transparency has been repeatedly identified as a key benefit of blockchain technology for the construction and engineering industry [48-51]. This value proposition is also appealing from the standpoint of practitioners [52]. Therefore, it has become a common objective for most existing studies. In reviewing the literature, this work employs a supply chain flow lens, categorizing the studies by their focus on the management of *information*, *product*, and *cash* flows.

*2.1.1. Information Flow*

Decentralized organizations are less resistant to the flow of information [53] and can benefit from increased transparency due to the higher degree of involvement from project participants [54]. This is crucial because the construction and engineering industry suffers from poor process transparency [55].

The consensus mechanisms underlying blockchains can empower decentralized models of data management, increasing trust in both the project information and the computational methods employed in project management. In managing the flow of information, the role of blockchain is not storing the data, but keeping an immutable record of changes to data [56]. This ability to track provenance is a key driver of increased transparency [57,58], providing independent parties with a quick means of verifying changes [59]. This is argued to foster collaboration among project stakeholders [60]. Furthermore, the application of blockchain can create a common information layer and promote the use of unified data definitions; the latter is crucial to project knowledge representation [61].

The blockchain technology can potentially benefit BIM workflows [62,63] which often involve independent parties working on a shared product model during the design, construction, and operation phases. Some argue that blockchain can enhance the privacy, data security, and permission management in BIM workflows [64]. A blockchain-based framework was proposed to address the security concerns for BIM-based workflows [65]. Others argue that the integration between BIM and blockchain can enable the decentralization of architectural design [66].

A framework was introduced for integrating BIM in the design and construction process using blockchain [67]; it focused on both the structure of interactions between project participants and the flow of information. Another work [68] stores the cryptographic hashes of BIMs on a blockchain to track changes. The log files generated by BIM-authoring solutions were categorized and used for recording the design events on a blockchain-based framework [69]. To streamline information sharing, a recent study used blockchain to authenticate data written to digital twins [70], enhancing confidence in the project data.

A limitation of existing studies is the data redundancy, with multiple copies of BIM stored off-chain. To address this limitation, two studies propose that only the changes to BIM need to be recorded rather than entire models. CryptoBIM [71] created a granular and content-addressable modeling of building information data on a blockchain, recording only the incremental changes to BIM elements and their semantics. A semantic differential transaction approach was introduced [72] to record the local model changes in the form of BIM contract changes. The approach was shown to significantly reduce the storage requirement for BIM workflows.

The research landscape on the management of information flow is not limited to BIM; the application of blockchain has been proposed for tracking the energy performance of buildings [73], estimating embodied carbon content [74], tracking quality information of prefabricated building elements [75,76], operation records for modular construction [77], data management in smart construction sites [78], and decentralizing the quality acceptance process [79].

A conceptual blockchain-based framework was proposed for managing asset information during operation [80]; it needs to be validated in future. It was argued that the integration between sensing platforms, smart contracts, and information models can enhance the tracking of maintenance and operation for physical assets [81]. A real-world case study application of blockchain in water infrastructure projects revealed that this technology reduces the effort associated with information management by 49% [82].

### 2.1.2. Product Flow

Enhancing the communications around material/product flow between suppliers, subcontractors, and general contractors can reduce the lead time, increasing service levels, and therefore reduce the construction cost [83]. The blockchain's traceability feature can facilitate this goal. One study [84] separated the material traceability into two components: tracing the origins of materials and tracking them through the supply chain. Based on a case study of a hypothetical supply chain, it was argued that the technology can be more effective with the latter objective [84]. For blockchain to be successful in product flow applications, it must be integrated with remote sensing and reality capture technologies [84,85].

To address this limitation, few studies focused on connecting the physical and digital worlds on construction projects: one study argued that the combined use of radio-frequency identification (RFID) and blockchain can improve the tracking of concrete panels from production to on-site delivery [86]; however, it remains to be validated. The integrated use of sensor recordings, BIM, and smart contracts was shown to be promising for tracking the installation of building façade panels [87]. A smart contract-based solution [88] created an immutable and content-addressable record of construction progress on a private Inter Planetary File System (IPFS) network [89] with immutable links to Ethereum blockchain [90,91]. These studies create a link between off- and on-chain realities, a crucial step toward the integration of product flow with the flow of information and funds.

### 2.1.3. Cash Flow

The industry's challenges with respect to cash flow management has motivated a move toward smart contracts and blockchain: 1) current payment applications and contracting practices are dependent on highly intermediated and manual workflows [92,93]; 2) the financial supply chain is highly fragmented and reliant on third-party institutions that hurt the visibility of cash flow [42,94]; 3) project participants have competing objectives, and therefore can exhibit opportunistic behaviors and provide incomplete or falsified information [95]. The resulting lack of transparency can lead to corruption [96].

Smart contracts can enhance transparency of cash flow by their formalization of rules and relationships [97,98], providing visibility regarding the performance of project participants prior and after they enter a contractual agreement. Automated processes can improve the financial supply chain [99], and smart contracts can add reliability to such automation by eliminating its single points of failure and centralized control mechanisms [100]. The blockchain technology, underlying smart contracts and decentralized applications, can be thought of as a "confidence machine" [101] that ensures the integrity and transparency of the underlying data used in payment processing [102]. This confidence stems from the use of decentralized consensus mechanisms such as Nakamoto consensus [103] and similar algorithms.

These potentials have created optimism with respect to the application of smart contract and blockchain for the management of project funds and contracting [104]. The increased transparency enabled by smart contracts is speculated to promote collaboration [105], reduce counterfeiting and fraudulent activity [106], and result in time and cost savings during both the project execution [107] and the contracting phase [108].

A framework [109] was proposed to enhance the transparency of the profit pool in integrated project delivery (IPD) projects, where planned and achieved profits are calculated automatically using a smart contract based on Hyperledger Fabric [110]. A conceptual business model [111] was developed for use by independent construction consultants and in the context of Swedish construction industry; the model aims to integrate the supply chain flows while providing transparency by its involvement of major project participants in its consensus mechanism. A case study of Nigeria reveals that the application of blockchain in public infrastructure projects creates transparency around government processes and reduces the possibility of corruption [112]. It was shown [113] that crypto assets (crypto currencies and crypto tokens) can be used to integrate a project's cash flow with its product flow, achieving higher granularity and atomicity compared with today's best practices.

A series of studies focused on providing transparency with respect to the financial status of project participants and the status of payments. A smart contract-based solution was developed to improve transparency around the clients' ability to pay [114]; it allowed for funds to be secured for 30 days, reducing the possibility of default. This can reduce the information asymmetry around the owners' financial status, a key challenge in both bidding and construction phase [7]. A cryptographic key management solution [115] provided "selective-transparency" between project stakeholders such that the sensitive information can be securely shared between two contracting parties. To enhance the transparency of the payment applications, a semi-autonomous consensus mechanism [116] was introduced; a smart contract executes payments once a peer-to-peer network of project members reach agreement with regard to the accuracy of the data provided by contractors in their applications for payment.

Others take a BIM-based approach toward enhancing transparency of payments. The integrated use of 5D BIM and smart contract was proposed for automated billing [117,118], where BIM use provided transparency with respect to the bill of quantities and project cost information used in the valuation of the work. The proposed framework will be implemented and validated in future. Two studies created visibility with regard to the status of cash flow in relationship with the status of product flow (e.g., off-site fabrication, delivery, on-site installation and construction), retrieving the latter directly from BIM: a smart contract-based framework [87] processed payments based on the delivery or installation of building façade panels; the approach was successfully tested in real-world, but the design of smart contract was not provided. Another work [88] used smart contracts to automate payments to subcontractors and general contractors based on the as-built status of projects documented by robotic reality capture. These studies have a potential to eliminate the need for intermediated payment applications and provide an integrated and transparent record of both product and cash flows.

### 2.1.4. Gap in Knowledge

For information management systems to be successful, their designers should think about how stakeholders use the information, and not just how they use the system [119]. It is necessary to go beyond the mere assessment of usability and to measure the impact of new technological solutions on information visibility via quantified value assessments [120]. Solutions based on blockchain and smart contract should not be an exception.

The existing body of research has clearly shown that these technologies have *potential* to enhance the visibility of information related to construction supply chain. Despite this growing enthusiasm and increased clarity regarding the know-how, we lack empirical analyses of how this technology compares with the industry's current best practices in their effectiveness and support of SCV.

Section 3 elaborates the two research questions explored in this work that aim to address this gap in knowledge.

## 3. Research Questions

This work explores the impact of smart contract and blockchain on SCV in the context of construction payments; this domain provides a great test bed for a few reasons: first, achieving transparency regarding payments necessitates a clear understanding of the two flows of cash and product, along with their integration. This is more difficult to achieve than analyzing each flow in isolation. Second, creating a coherent and singular documentation of these supply chain flows is a challenge due to the industry's reliance on external financial institutions in its already fragmented supply chain.

The SCV definition used in this work is borrowed from a review of SCV literature [121]: "supply chain visibility is the identity, location and status of entities transiting the supply chain, captured in timely messages about events, along with the planned and actual dates/times for these events." In the context of construction payments, these entities comprise the flows of cash and product.

When analyzing the visibility of construction payments, it is crucial to take *granularity* into consideration; it directly affects the complexity of information retrieval [113]. Payments can be distinguished by their 1) *temporal*, 2) *trade*-level, and 3) *product*-level granularity [113]: these respectively correspond to the period for which a trade is compensated, the number of trades compensated in a payment, and the scope of valued construction work. Therefore, in its impact assessment, this work explores the following two research questions (RQ):

*RQ1:* How does the visibility of the conventional and blockchain-enabled construction payment systems compare at different levels of granularity?

*RQ2:* How does increased granularity (time, trade, product) affect visibility in each payment system?

Fig. 1 shows the comparisons that are drawn for each research question.

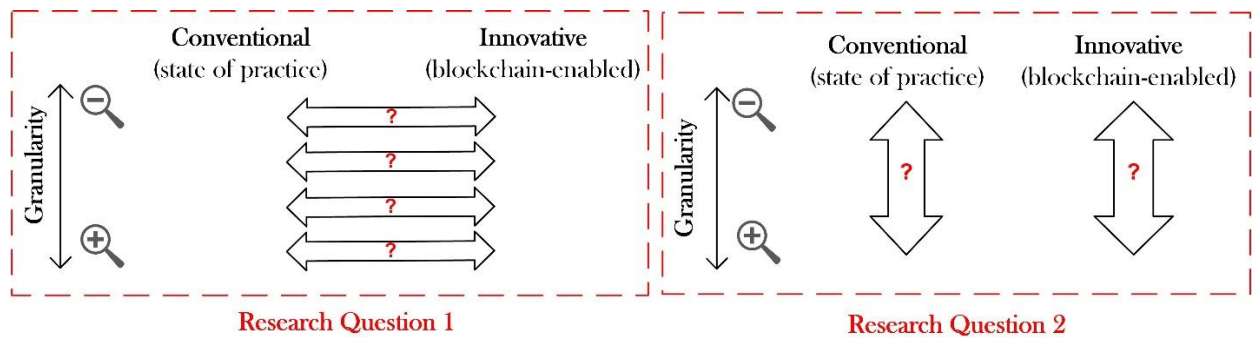

Fig. 1. The research questions respectively 1) draw comparison between the two processes at various levels of granularity, and 2) study the performance of each individual process with respect to increased granularity

## 4. Method: Charrette Test

The Charrette Test Method [122] was employed to answer these two questions. It uses comparative empirical experiments to evaluate the "effectiveness" of different tools in performing similar tasks. It is commonly used for rigorous validation of construction research [123]. Herein a comparison is drawn between 1) a payment system based on blockchain-enabled smart contracts, and 2) one based on the industry's state-of-practice payment systems. These are respectively referred to as "*innovative process*" and "*conventional process*" in the Charrette Test Method.

A total of 14 participants, comprising of doctoral candidates in the construction engineering and management program at Stanford University (USA), were assigned to either one of the two payment systems. All participants have proficiency in general construction management concepts. In each trial, a participant completed a number of tasks: they interacted with the assigned payment system to determine the amounts of payments made to trades under scenarios with varying levels of granularity.

Prior to performing the tasks, participants watched an 8-minute instructional video and spent 5 minutes familiarizing themselves with the assigned payment system. The instructional videos were produced separately for the innovative and conventional processes. In both, the authors introduced payment workflows and demonstrated a few examples of how tasks can be completed with a given payment system.

This resulted in 14 trials (7 for each process) which were evaluated using a series of metrics to arrive at a comparison of how the two processes support SCV. The following sections respectively elaborate on the design of blockchain-enabled and conventional *processes* used by Charrette test participants (section 4.1), the design of the *tasks* completed by participants (section 4.2), and the *metrics* used in performance evaluation (section 4.3). The charrette test results are provided in section 5.

### *4.1. Innovative and Conventional Processes*

Both processes were modeled in the form of interactive dashboards and implemented in Tableau software. Both dashboards display the payments corresponding to three months of indoor finishing work by four subcontractors at a commercial construction project in Toronto (Canada). The 3,400 $m^2$ job site was inspected using a camera-equipped unmanned aerial vehicle; the robot-captured visual data was passed to a computer vision-based algorithm [124,125] to arrive at the state of progress for indoor partitions: percentages of completion were reported for framing, insulation, drywalling, plastering, and painting.

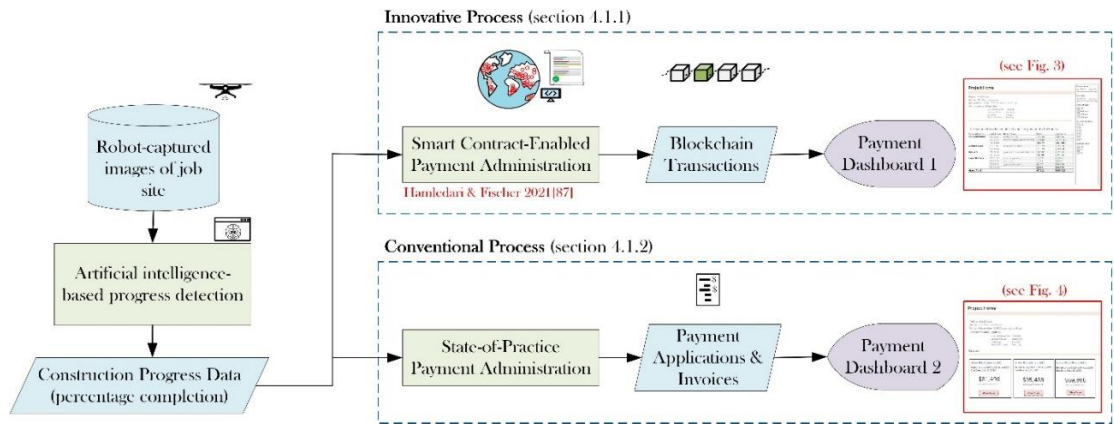

Fig. 2. Progress data captured at a commercial construction site was used to generate the payment data for both innovative (blockchain-enabled) and conventional (state-of-practice) processes

This progress data was passed to *both* the innovative and conventional processes (Fig. 2). The two separately arrived at the payments for the months of June, July, and August; their corresponding dashboards visualize the *same* total amounts of payments ($247k) to the *same* four subcontractors.

### 4.1.1. Blockchain-Enabled Payment Administration

In the innovative process, a smart contract-enabled method [88], developed by the authors, was used to pay the four trades. This solution stores the progress data off-chain on a private IPFS network and communicates the valuations to a smart contract via a JSON remote procedure call. The smart contract, running on the Ethereum Virtual Machine, performs on-chain payment settlements and the transfers of lien rights respectively using the instances of crypto currencies and ERC-721 non-fungible crypto tokens. The resulting 406 transactions were written to the Ethereum blockchain and used as input to the interactive dashboard (Fig. 3) (please consult [88,100] for the design of the smart contract algorithms).

The dashboard used by participants in the Charrette test is illustrated in Fig. 3: the payments to the four trades are listed based on cost code, and a series of filters can be used to narrow down the scope of visualized payments (e.g., based on when the work was performed, when the payments were made, the element types, and the building floors where the work was performed). The dashboard is based on the 406 transactions sitting on the blockchain; an example of transactions processed by the smart contract is shown in Fig. 3b

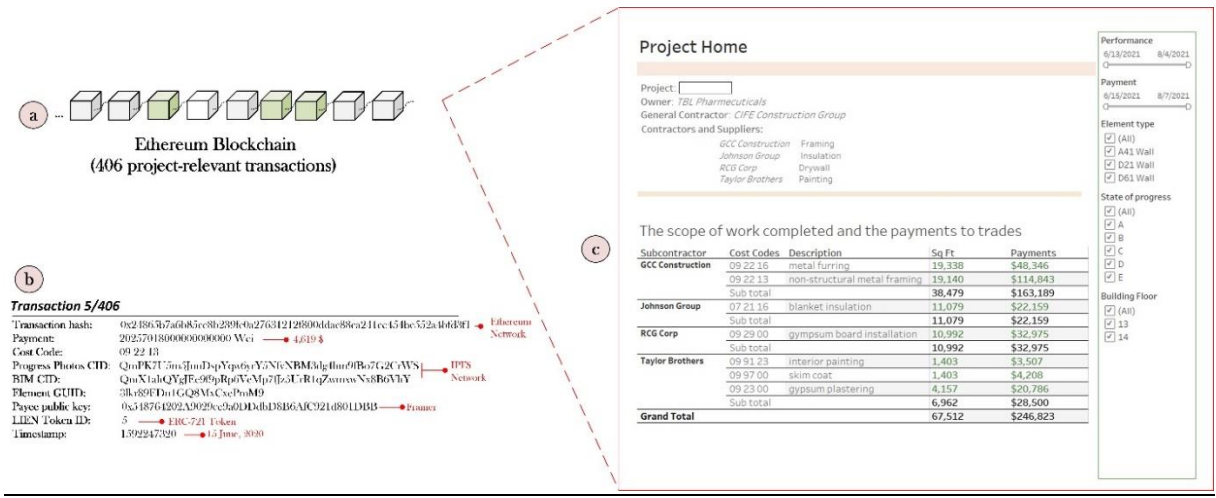

Fig. 3. The dashboard displaying the payment data for the innovative process: a) smart contract-enabled payments written to the Ethereum blockchain, b) an example of a transaction compensating the framer, and c) the interactive dashboard based on the 406 transactions; the dashboard was used by the participants in the Charrette test

*4.1.2. State-of-Practice Payment System*

In the conventional process, general contractors (GC) prepare monthly applications for payments using standardized forms (e.g., AIA G702 and G703) where all scopes of work completed by trades are listed and submitted to owner for review. The industry still heavily relies on paper-based workflows but has recently started a move toward digital alternatives.

To arrive at an accurate definition of *state-of-practice*, the authors conducted interviews with two major US-based construction firms with leading digital initiatives; two internet-based software applications were identified as the primary technological solutions used in the industry, both of which digitalize the workflows for the preparation, documentation, and visualization of valued work listed in AIA G702 and G703. The dashboard used in the conventional process (Fig. 4) was closely modeled after these two software applications to establish a fair representation of state of practice. The dashboard displays (Fig. 4) the three payment applications (draws) for the months of June ($82k), July ($95k), and August ($70k). Please note that these payments amount to the same total as the smart contract-based solution (247k) described in section 4.1.1.

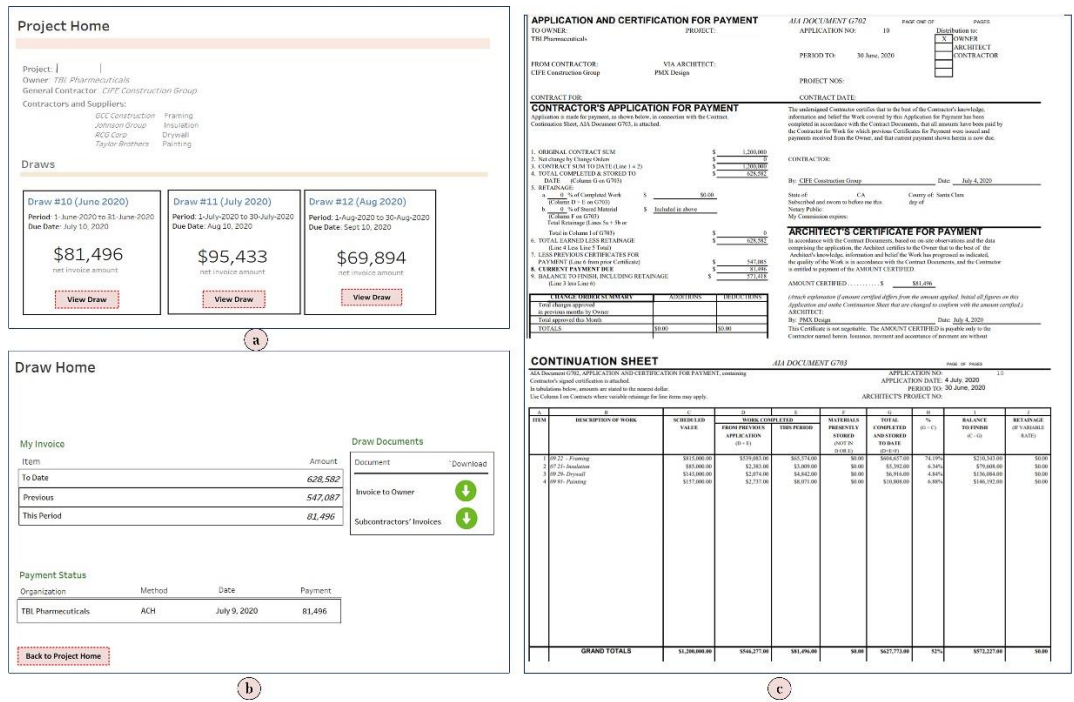

Fig. 4. The conventional process: a) the payment dashboard used by participants in the Charrette test, b) the draw home providing a summary of payments for that month along with supporting AIA documents, and c) an example of the application for payment G702 and G703 documenting payments based on the cost code, valuation, and the description of the work

Participants can view each draw page (Fig. 4b) to review the supporting documents including the invoices submitted to the GC by the four trades and the overall invoice submitted to the owner by GC. Each invoice breaks down payments in terms of cost codes, their valuations, and the description of the work (Fig. 4c). These invoices were prepared by the authors and based on the input progress data (Fig. 2).

## *4.2. Tasks*

In each trial, a participant is asked to complete 12 tasks using the provided dashboard. Each task asks the participant to identify the amount of payments to trades under a scenario. As previously explained, these scenarios are distinguished by their choices of *temporal*, *trade-level*, and *product-level* factors. Each factor can be *low (L)* or *high* (H) in its granularity (Table 1). For example, inquiring about the "payments made for painting" and "payments made for work on the 14th floor partitions" respectively have low and high *product-level granularity* according to Table 1. The latter describes specific instances of building elements (product).

Table 1. The levels of granularity used for temporal, trade-level, and product-level factors

| Factor | Granularity | Description |
|---|---|---|
| Time | High | one week |
| | Low | one month |
| Trade | High | direct payment to a subcontractor |
| | Low | payments to GC/few subcontractors |
| Product | High | referring to the elements' location, type, or cost code |
| | Low | describing general scopes of work (e.g., plastering) |

Depending on the choice of three factors, each task can have one of the *four* levels of granularity: *lowest* (3 L), *low* (2 L, 1 H), *high* (2 H, 1 L), and *highest* (3 H). Three tasks were designed for each level of granularity, resulting in a total of 12 tasks that will be completed by each participant; they are listed in Table 2. Participants either provide a written response or skip the task if they are unable to find an answer.

Table 2. The 12 tasks performed by participants in the Charrette Test Method

| Task | Granularity | Trade, Time, Product | Description |
|---|---|---|---|
| 1 | lowest | LLL | Payments to the General Contractor (GC) made in July |
| 2 | lowest | LLL | Payments to GC for the insulation work performed in June |
| 3 | lowest | LLL | Payments to GC for plastering and painting performed in August |
| 4 | low | LLH | Payments to GC for the work on the 14th floor's D21 walls performed in July |
| 5 | low | LHL | Payments to GC between July 15-July 22 |
| 6 | low | HLL | Payments to RCG Corp* for work performed June 15- July 15 |
| 7 | high | LHH | Payments to GC for the work performed on A41 walls between July 8- July 15 |
| 8 | high | HLH | Payments to GCC Construction made between June 15-July 15 for Metal Furring (09 22 16) |
| 9 | high | HHL | Payments to Taylor Brothers made between June 8-July 15 |
| 10 | highest | HHH | Payments to Johnson Group made between Aug 1-Aug 7 for Blanket Insulation (07 21 16) on 13th floor |
| 11 | highest | HHH | Payments to Taylor Brothers for Gypsum Plastering (09 23 00) on D21 walls performed July 28-Aug 4 |
| 12 | highest | HHH | Payments to RCG Corp for the work on 14th floor performed June 15-June 22 |

* The four subcontractors in the worked example include: GCC construction (framing), Johnson Group (insulation), RCG Corp (drywalling), Taylor Brothers (painting)

## 4.3. SCV Metrics

Three metrics were defined to evaluate the performance of participants. These were adapted from a recent systematic review of SCV literature [126] which categorized SCV metrics into informational, automational, and transformational; the former was more prominently used in the literature and concerns the quality of information collected and distributed among supply chain partners.

With respect to informational characteristics of SCV, the most commonly used metrics include *accuracy*, *completeness*, and *timeliness* [126]. Therefore, this work defines the following metrics for the Charrette Test Method:

1) *Information accuracy:* a task is accurately completed, if the participant provides a written response which matches the ground truth (i.e., the actual amount of payment under that scenario). Rounding up and down was allowed.
2) *Information completeness:* a task is considered complete, if the participant provides a written response (regardless of its accuracy) and does not skip the task.
3) *Information latency*: it measures the time spent on completed tasks (in seconds).

# 5. Results & Discussions

To address the RQ1, the results of the 14 trials were used to compare the three SCV metrics between the blockchain-enabled and conventional processes at different levels of granularity (section 5.1). To address RQ2, the impact of increased granularity on the SCV metrics was independently evaluated for each of the two processes (section 5.2). These comparisons are illustrated in Fig. 1. The independent-samples, two-tailed t-test was used to evaluate the statistical significance of findings. P values greater than 0.05 ($P>0.05$) were considered "not significant".

## 5.1. Comparison of Blockchain-Enabled and Conventional Processes (RQ1)

To compare the two processes at the four levels of granularity, the SCV metrics were independently calculated based on participants' responses to tasks 1-3 (lowest granularity), tasks 4-6 (low granularity), tasks 7-9 (high granularity), and tasks 10-12 (highest granularity):

At the *lowest granularity* (tasks 1-3), blockchain-enabled process provided no significant improvement for any of the three SCV metrics (Table 3): the blockchain-enabled process improved information completeness by 5%, achieved the same accuracy, and had 11% higher latency; none of these, however, were observed to be significant (Table 3).

Table 3. Comparison of the two processes at the lowest granularity (tasks 1-3)

|  | Blockchain-Enabled | | Conventional | | T-test (df=40) |
| --- | --- | --- | --- | --- | --- |
|  | Average | St.dev | Average | St.dev |  |
| Completeness (%) | 100 | 0 | 95.24 | 21.82 | not significant (*p=0.33*) |
| Accuracy (%) | 85.71 | 35.82 | 85.71 | 35.86 | not significant (*p=1.00*) |
| Latency (seconds) | 83.10 | 58.34 | 75.00 | 22.07 | not significant (*p=0.56*) |

Blockchain-enabled process shined when at least one of the trade, time, and product factors were high in granularity (tasks 4-12): the blockchain-enabled process increased the *information completeness* by 3 times, 2.6 times, and 6.4 times respectively at low, high, and highest granularity. As shown in Fig.5, this performance improvement was observed to be statistically significant at all three levels of low (df=40, p=4E-6), high (df=40, p=2E-5), and highest granularity (df=40, p=6E-9).

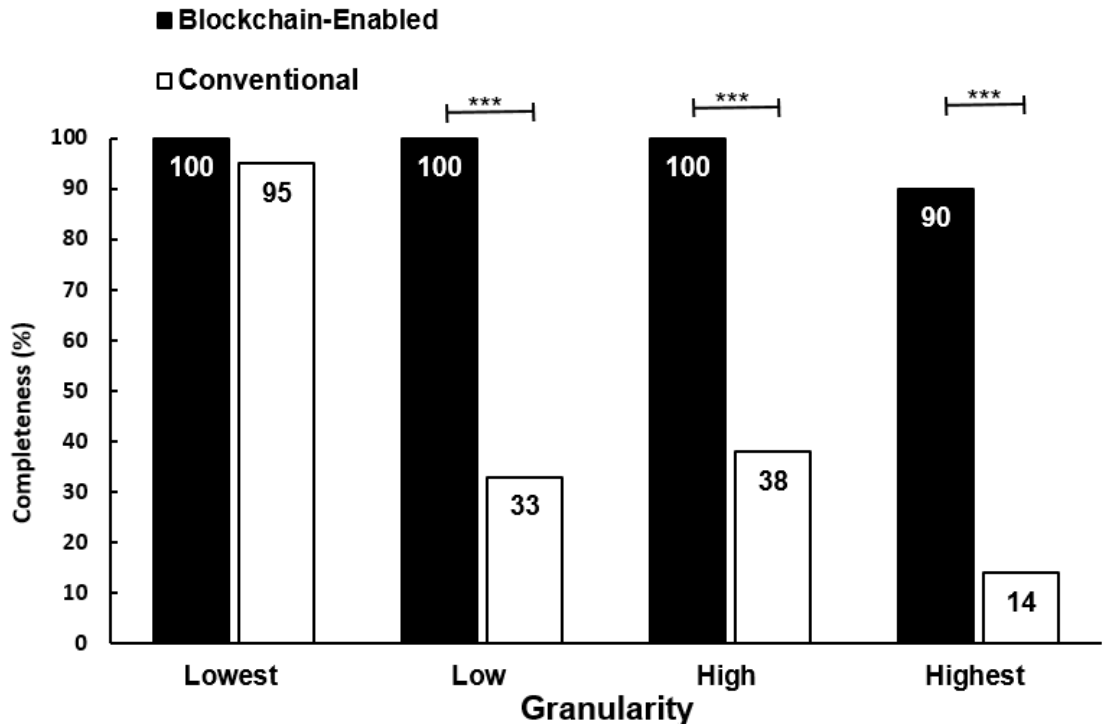

Fig. 5. Information completeness scores across the four levels of granularity

Similar observations were made in terms of *information accuracy*: the blockchain-enabled process increased this metric in all levels of granularity except the lowest (Fig. 6). The performance improvement was significant at the low (df=40, p=3E-10), high (df=40, p=3E-4), and highest granularity (df=40, p=3E-10).

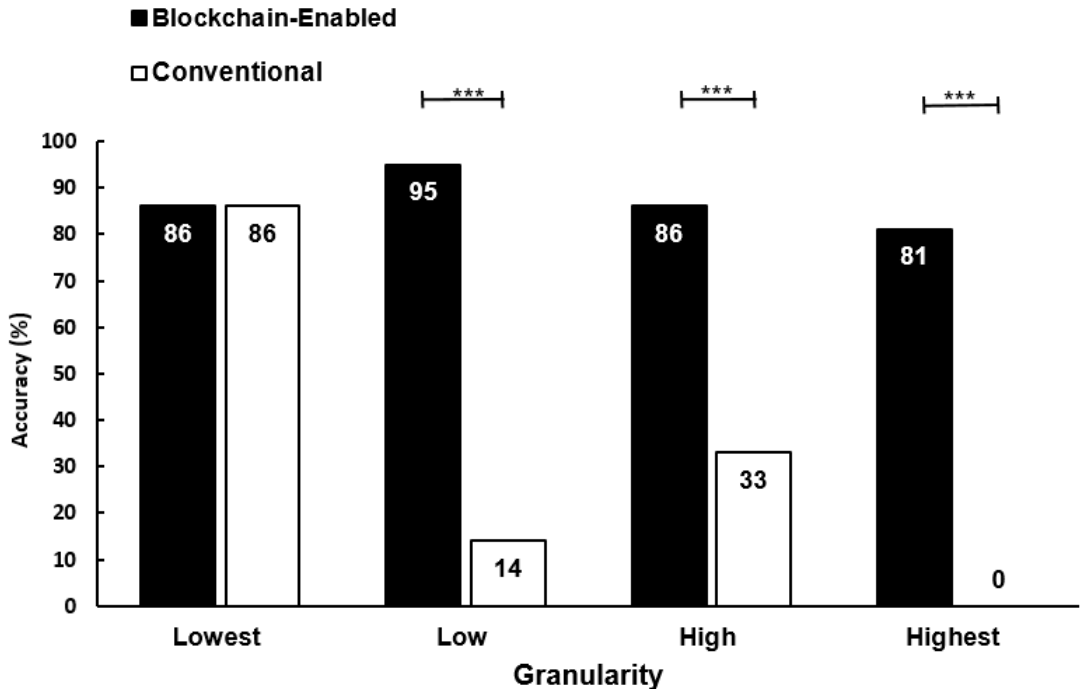

Fig. 6. Information accuracy scores across the four levels of granularity

Unlike the other two metrics, the *information latency* was not improved by the blockchain-enabled process at any level of granularity (Fig. 7): at the lowest granularity, the participants using the conventional process were 10% faster. This difference, however, was not statistically significant (df=39, p=0.56).

In the other three levels of granularity (low, high, and highest), the relationship was reversed: participants using the blockchain-enabled process were faster in retrieving information (i.e., lower information latency). The results were neither statistically significant, nor they appeared reliable due to the large inequality of sample sizes between the two processes: the number of observations were considerably lower for the conventional process at those three levels of granularity. This is because the information latency is only measured for completed tasks; in the conventional process, the information completeness dropped to 33%, 38%, and 14% respectively for low, high, and highest granularity (Fig. 5). Therefore, no conclusion was reached for the impact of blockchain on information latency at the low, high, and highest granularity.

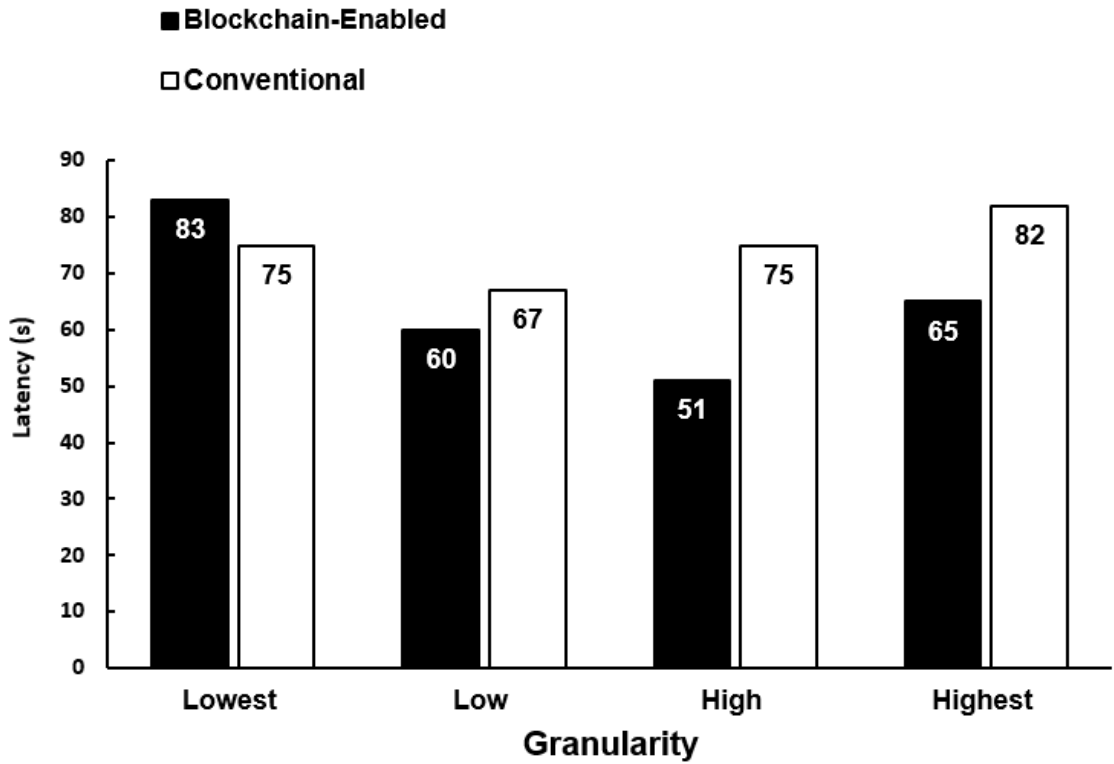

Fig. 7. Information latency across the four levels of granularity

## *5.2. Effect of Increased Granularity (RQ2)*

Each of the two processes was independently studied to assess its support of increased granularity. The level of granularity can be increased in 6 potential scenarios (combinations of 2 out of 4 levels of granularity). For each scenario (e.g., transition from lowest to high), the independent-samples, two-tailed t-test was used to evaluate whether the two SCV metrics of information accuracy and completeness degrade due to increased granularity.

For example, to assess whether the conventional process's accuracy degrades in the transition from the lowest to high granularity, the participants' accuracy at the lowest granularity (tasks 1-3) was compared with their accuracy at the high granularity (tasks 7-9) using the t-test (df=40).

The blockchain-enabled process proved more robust in supporting increased granularity compared with the conventional one:

- In the blockchain-enabled process, increased granularity *never* caused degradation in either the completeness or the accuracy of the information.
- In the conventional process, on the other hand, the information completeness is *50% likely* (3/6) to degrade when the granularity is increased.
- In the conventional process, the information accuracy is *66.7% likely* (4/6) to degrade when granularity is increased.

## *5.3. Implications*

To ensure future use cases of blockchain and smart contract can provide similar SCV benefits, it is imperative for construction and engineering firms to 1) focus on the underlying mechanisms that are key to blockchain-enabled visibility (section 5.3.1) and emulate them in their own implementations of blockchain; and 2) understand the business impact of increased supply chain visibility (section 5.3.2) to guide their decision making with regard to the degree of adoption.

### *5.3.1. Underlying Mechanisms of Blockchain-Enabled Visibility*

The *integration* between the cash flow and the product flow is a key factor to achieving visibility. In the conventional payment process, this integration is weak: Each payment application provides an *aggregate* view of all work on the job site during a one-month period. In the blockchain-enabled process, on the other hand, the smart contract can directly condition the flow of cash based on the on-site observations of product flow.

This blockchain-enabled integration of product and cash flows, as achieved by the implementation used in this work, can be distinguished from that achieved by today's conventional process in two ways [113]: 1) it is *atomic*: the two flows are not documented separately but they are stored together. Having access to either guarantees access to the other: studying the transactions on the Ethereum blockchain, project members can directly query the scopes of work and reality captures that triggered payments and vice versa; 2) it can be *granular*: the smart contract enables project members to narrow the scope of payments or the scope of their analysis in terms of the number of parties compensated in a payment (trade), the period for which work is compensated (time), and the number of building elements (product).

The atomicity and the granularity of supply chain flow integration is key to achieving visibility. In addition, it results in semantically rich documentations of cash and product flows. Section 5.3.2 discusses how firms can use such data to better understand the financial performance of their projects.

### *5.3.2. Value Creation: Enhancing the Economic Profit*

The construction and engineering industry has been notorious for its poor productivity, a problem well documented by both academics [127,128] and industry practitioners [129,130]. Enhancing value is therefore critical to the firms operating in this sector.

Value is created only if the return on invested capital (ROIC) exceeds the opportunity cost of capital [131]. To improve their *economic* profit, firms need to monitor the cost structure of projects and pinpoint areas for improvement by analyzing the invested capital, the resulting returns, and opportunity costs. This analysis is difficult to perform at a *granular* level because construction and engineering firms lack such detailed documentation of cash and product flows. They instead have a broad view as provided by today's payment systems.

This is where blockchain-enabled visibility can play a huge role. It provides firms with a rich set of data that is both atomic and granular. This is a powerful tool at the disposal of firms that need to take a deeper dive into their financial data. Firms can use such data to slice the cost and revenue across one or a combination of factors such as, but not limited to, product lines, time periods, and trades involved in a project. Such analyses of financial data can help identify areas that suffer from poor productivity and low returns on invested capital.

For example, a general contractor self-performing different scopes of work can benefit from the blockchain-enabled visibility by breaking down project cost and revenue across the self-performed building elements to draw comparisons, identify areas of low return, and make strategic decisions to either remedy the problem of poor productivity or seek alternative solutions that may provide better ROIC (e.g., outsourcing). Similar analyses can be conducted across periods of time or across trades. The latter, for example, can enable firms to factor in historical performances as they decide who they partner with on a new project. This can particularly benefit participants in projects with a shared reward structure (e.g., integrated project delivery).

## 6. Limitations and Future Work

The following concerns and limitations must be taken into consideration when interpreting the results presented in this work:

### *6.1. The Perils of Overgeneralization*

While the application of blockchain and smart contracts improved the construction supply chain visibility, the authors strongly caution readers against overgeneralization of findings: the charrette test method validates the reliability and generalizability for the effectiveness of the particular blockchain-enabled method tested in this work. This does show promise for the underlying technology. However, the observed performance improvements in information completeness and accuracy will not necessarily extend to other implementations.

The construction and engineering firms can see this work's results as a positive sign that blockchain and smart contract, when used correctly, indeed have potential for enhancing visibility. It is necessary for firms to proactively 1) define their business needs that motivate the use of smart contract and blockchain, 2) use those objectives to define metrics of success, and 3) test the effectiveness of developed prototypes in real-world scenarios.

For example, firms can strategize their adoption of blockchain based on their needs and this work's findings: those that only need a broader view of cash and product flows are not likely to enjoy high returns on their investments in the blockchain space (when return is measured in terms of information completeness and accuracy); they can enjoy equal returns using conventional platforms and with lower capital investments.

### *6.2. Connecting Off- and On-Chain Realities*

The blockchain-enabled process outperformed the conventional one in terms of both information completeness and information accuracy. However, the latter metric (accuracy) requires closer scrutiny:

The charrette test method used the valuation of the work and the payments as ground truth for measuring the accuracy of participants' responses. For both processes, these valuations were machine-generated and based on a series of robotic reality captures (Fig. 2). These progress reports, whether generated by machines or humans, are themselves prone to inaccuracy. For example, there can be discrepancies between the *actual* and the *reported* percentage completion of work, the amount of material delivered on site, and the quality of construction work.

This motivates a close look into the design of robust 'oracles', solutions that provide blockchain and smart contracts with off-chain realities (the status of construction and engineering projects). If based on poorly designed oracles, blockchain-enabled information systems provide visibility into a misrepresented view of job sites.

The blockchain implementation presented in this work used camera-mounted unmanned aerial vehicles and artificial intelligence as its oracle. These solutions were independently tested for their accuracy before integration with smart contract.

### *6.3. More Comprehensive View of SCV*

There are three aspects to SCV: *informational*, *automational*, and *transformational* [126]. This work studied the impact of the technology on the former. To arrive at a more comprehensive view of SCV, future work should focus on impact assessment for the other two aspects:

The *automational* characteristics of SCV are commonly measured in terms of automated information capturing and automated information integration [126]. Blockchain-enabled solutions are shown to enhance the quality of integration between supply chain flows [113]. However, it is not clear 1) whether such enhancements translate to superior performance for the two metrics discussed above and 2) whether these metrics can fully capture the value added by the technology. The state-of-the-practice digital platforms can be equally capable of achieving *automation* without reliance on blockchain. It has been argued that *reliable automation*, and not the mere automation of workflows, is the true value added by blockchain-enabled smart contracts [100]. This is the key distinction that separates blockchain's automational characteristics from that of today's solutions, and it may necessitate the definition of new metrics that can better capture these nuances in different modes of automation.

The *transformational* characteristics of SCV are concerned with how information is utilized by firms to enhance their competitive edge or improve their operational efficiency [126]. While more difficult to measure, understanding this perspective is paramount to successful adoption of blockchain and smart contract. The business transformations fueled by blockchain are still in early stages [132], and much more work is needed in connecting business objectives with technology's core capabilities.

# 7. Conclusions

Blockchain and smart contract technologies have a long way to go before they earn their over-hyped status. They have been touted for their 'potential' impact on construction supply chain visibility. However, a scarcity of real-world implementations coupled with a lack of impact assessments have created a sentiment toward these technologies that is a mix of enthusiasm and pessimism.

To address this gap, the paper used comparative empirical experiments (Charrette Test Method) to measure the impact of blockchain and smart contract on the visibility of construction supply chain and in the context of payments (the intersection of cash flow and product flow). In a series of real-world trials, a conventional payment system was contrasted with one based on blockchain-enabled smart contracts: the two were compared with respect to their support of 12 information retrieval tasks with four levels of granularity. The three metrics of information completeness, information accuracy, and information latency were used to quantify the performance of charrette test participants. The application of blockchain was found to enhance the first two metrics as discussed below.

The findings can guide construction and engineering firms as they navigate the hype around these technologies, identify the right business opportunities, and make strategic decisions regarding the degree of technology adoption:

First, blockchain-enabled solutions are expected to provide superior supply chain visibility only for firms that need a more granular look into the flows of cash and product. For inquiries associated with the three higher levels of granularity, blockchain-enabled solution enhanced information completeness by 341% compared with the conventional solution; the information accuracy was enhanced by 557%. Participants needing a broad view (tasks with the lowest granularity) saw no visibility benefits by using the blockchain-enabled solution.

Second, firms adopting blockchain can enjoy higher returns on their invested capital by increasing the granularity of the lens they use for analyzing their financial performance and their product flow management. According to the charrette test results, the blockchain-enabled solution provided no additional visibility benefits for inquiries that have the lowest granularity. To use blockchain-enabled solutions to their full potential, firms must increase the granularity of the temporal, product-level, and trade-level scope of their analysis. The blockchain-enabled solution was supportive of this increased granularity; its information completeness and accuracy did not suffer. Faced with the same increased granularity, firms using the state-of-practice solutions are 50% and 66.7% likely to experience decline respectively in the completeness and accuracy of information.

The results emphasize the importance of supply chain information systems that emulate visibility *by design*. While the conventional system was on average 45% less accurate than the blockchain-enabled one, participants using the conventional and the blockchain-enabled process were *equally confident* when asked about the accuracy of their responses (df=12, p=0.84). This has huge business implications for an industry whose practitioners spend most of their time managing information. Information systems must support visibility by design, and blockchain and smart contract provide promise due to their atomic and granular integration of supply chain flows, a key factor contributing to visibility.

While this work measured the impact of blockchain and smart contract on the informational aspects of construction supply chain visibility, it is equally crucial for future work to study the automational and transformational aspects of SCV to arrive at a more comprehensive view.

## 8. Acknowledgement

This work is financially supported by the Center for Integrated Facility Engineering (CIFE) at Stanford University (grants 2020-09, 2018-06, 2017-06). The authors are grateful to Swinerton, PMX Construction, Perkins+Will, Inc., for their support during the data collection phase and granting access to project data. The first author extends his gratitude to the following individuals for their invaluable intellectual companionship, their genuine friendship, and their immense role in his PhD journey: Eric Law and Tristen Magallanes (Swinerton); Dr. Kincho Law, Dr. Michael Lepech, Dr. Forest Flager, Howard W. Ashcraft, Ashwin Agrawal, Alissa Cooperman, Parisa Nikkhoo, Tulika Majumdar, and Yujin Lee (Stanford University); Dr. Yuting Chen (UNC Charlotte); Dr. Brenda McCabe, and Pouya Zangeneh (University of Toronto).